%% file: main.tex
\documentclass[conference,10pt,a4paper,twocolumns,final]{IEEEtran}

\newlength{\flexwidth}
\usepackage{amsmath,amssymb,amsthm,fixmath}
\usepackage{mathtools}
\usepackage{accents}
\usepackage{color,colortbl}
\usepackage{xcolor}
\usepackage{siunitx}
\usepackage{cuted}
\usepackage{multirow,booktabs}
\usepackage{subcaption}
\usepackage[inline]{enumitem}
\usepackage{optidef}
\usepackage{graphicx}
\usepackage{lipsum}

\usepackage{epstopdf}
\usepackage[textsize=tiny,colorinlistoftodos]{todonotes}

\usepackage[shortcuts,acronym,automake]{glossaries}
\input{glossary}

\newcommand{\subscript}[1]{_{\mathrm{#1}}}

\usepackage{tcolorbox}
\tcbuselibrary{many}

\newtcbtheorem[]{alg}{Algorithm}{fonttitle=\bfseries}{alg}
\usepackage[norelsize,linesnumbered,ruled]{algorithm2e}
\SetKwRepeat{Do}{do}{while}
\makeatletter
\newcommand{\removelatexerror} {\let\@latex@error\@gobble}
\makeatother

\begin{document}

	\title{A Secure and Robust Approach for Distance-Based Mutual Positioning of Unmanned Aerial Vehicles}
	
	\author{
		\IEEEauthorblockN{Bin~Han\IEEEauthorrefmark{1}~and~Hans~D.~Schotten\IEEEauthorrefmark{1}\IEEEauthorrefmark{2}}
		\IEEEauthorblockA{
		\IEEEauthorrefmark{1}Rheinland-Pf\"alzische Technische Universit\"at Kaiserslautern-Landau  (RPTU), Kaiserslautern, Germany\\
		\IEEEauthorrefmark{2}German Research Center of Artificial Intelligence (DFKI), Kaiserslautern, Germany
		}
	}
	
	\bstctlcite{IEEEexample:BSTcontrol}
	
	\maketitle
	
	\begin{abstract}
		\Ac{uav} is becoming increasingly important in modern civilian and military applications. However, its novel use cases is bottlenecked by conventional satellite and terrestrial localization technologies, and calling for complementary solutions. Multi-\ac{uav} mutual positioning can be a potential answer, but its accuracy and security are challenged by inaccurate and/or malicious measurements. This paper proposes a novel, robust, and secure approach to address these issues.
	\end{abstract}
	
	\begin{IEEEkeywords}
		\ac{uav}, mutual positioning, security, robustness
	\end{IEEEkeywords}
	
	\IEEEpeerreviewmaketitle
	
	\glsresetall
	\section{Introduction}\label{sec:introduction}
	Over the past decade, the importance of \ac{uav} has risen dramatically. Its deployment, especially in form of multi-\ac{uav} systems, has been rapidly growing in various civilian~\cite{SSA+2019unmanned} and military~\cite{SAB+2019design} use scenarios. Particularly, multi-\ac{uav} systems are widely considered to play an indispensable role in future \ac{6g}~\cite{JHH+2021road} and \ac{mec}~\cite{LCG+2022multi} systems.
	
	For safe and accurate operation, as well as efficient task execution, precise localization of the \glspl{uav} is commonly crucial in such systems. While satellite positioning functionalities, such as \ac{gps}, are usually available in conventional \ac{uav} devices nowadays, their accuracy may be affected in some particular environments. Such environments are including but not limited to dense urban area, indoor space, tunnels, dense foliage, and deep valleys. Unfortunately, these are also typical deployment scenarios of some important \ac{uav} use cases, such as emergency rescue, aerial mobile base station, geological prospecting, and environment monitoring. Other positioning methods are therefore required, to complement or replace the satellite solution in such situation.
	
	One possible answer is provided by the well developed terrestrial localization systems. These technologies are particularly feasible and reliable for consistent indoor deployment scenarios, generally relying on geolocation anchors fixed at known positions, such as landmark sensors~\cite{AMH2016rssi}, WiFi access points~\cite{YS2015wifi}, or cellular base stations~\cite{DSM+2021positioning}. However, for emergency deployment or remote areas, it can be technically unfeasible or cost-prohibitive to install sensors or WiFi devices, and the mobile networks can be hardly accessible, either.
	
	As an alternative approach, in this paper we investigate the mutual positioning, a.k.a. cross-positioning of multiple \glspl{uav}, which exploits the spatial diversity of different devices to achieve accurate localization. Like all other localization systems, multi-agent mutual positioning can be realized either direction-based or distance-based~\cite{KZS+2020review}. While the few existing research on this topic are mostly assuming direction information to be available~\cite{MMS+2013navigation,WWS+2020auxiliary}, in this study we focus on the distance-based option due to its low implementation complexity and less sensitivity to the line-of-sight constraints.
	
	While multi-\ac{uav} mutual positioning is following the same geometrical principle as other approaches, it is challenged by the lack of reliable landmarks. The positions of reference \glspl{uav}, unlike those of satellites or terrestrial anchors, are neither static nor accurate. Moreover, in complex radio environments, the mobility of \ac{uav} can generate strong channel fading effect, leading to high dynamics in the variance of distance measurement. These challenges are calling for a novel positioning model that considers inaccurate reference position and random power of distance error, which have not yet been studied to the best of our knowledge.
	 
	Moreover, like all other radio positioning approaches, mutual positioning is exposed to the risk of attacks, e.g. jamming or measurement falsification. This security problem has been well investigated by literature in the context of terrestrial positioning, where various approaches are proposed to defend and detect malicious data. However, tailored for sensor-based localization systems, conventional solutions generally rely on fixed landmarks \cite{WB2019robust} and/or deterministic variance across \glspl{uav} \cite{GVW2012efficient, MSK2021rss}, making them infeasible for our problem.
	
	In this paper, we present a novel secure framework for multi-\ac{uav} mutual positioning, including a \ac{rgd} algorithm that localizes the target \ac{uav} from inaccurate reference positions associated with distances measured under various variances, and a confidence-based \ac{rdad} frame that detects malicious data. The remainder of the paper is organized as follows: we start with Sec.~\ref{sec:problem_setup} outlining the system and attack models, then in Sec.~\ref{sec:robust_positioning} we introduce conventional position estimators, and develop an advanced error model to evolve them into our \ac{rgd}. Based on that, we propose in Sec.~\ref{sec:rdad} the \ac{rdad} scheme to detect malicious attacks and further enhance the robustness of the system. In the end, we demonstrate the effectiveness of our proposals with numerical results in Sec.~\ref{sec:simulations}, before concluding the work with Sec.~\ref{sec:conclusion}.
	
%
	\section{Problem Setup}\label{sec:problem_setup}
	\subsection{System Model}
\begin{figure}[!hbpt]
		\centering
		\includegraphics[width=.8\linewidth]{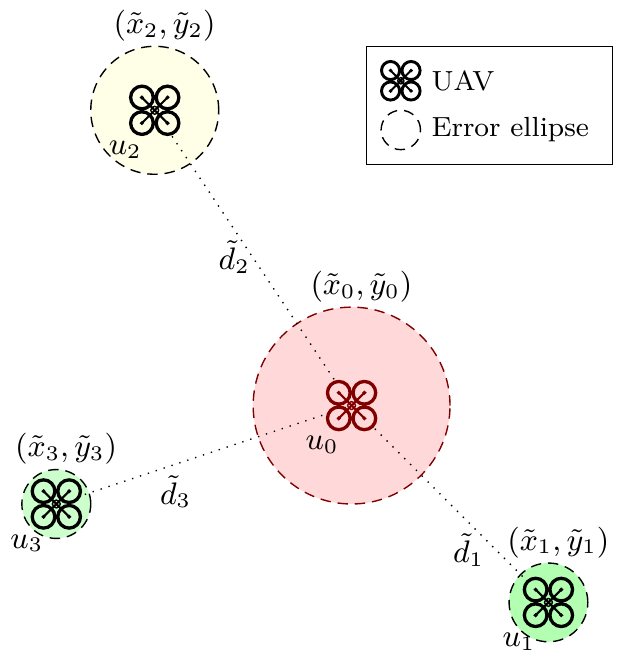}
		\caption{Concept of the multi-agent mutual positioning}
		\label{fig:concept}
	\end{figure}

	We consider a set of \glspl{uav} $\mathcal{U}=\{u_0,u_1,\dots u_I\}$, each equipped with a wireless communication module and a self-positioning module (e.g., \ac{gps}). At any given time $t$, every $u_i\in\mathcal{U}$ can measure its own position $(x_i,y_i)$ inaccurately:
	\begin{equation}
		\left[\tilde{x}_i(t),\tilde{y}_i(t)\right]=\left[x_i(t)+\delta_{x_i}(t),y_i(t)+\delta_{y_i}(t)\right],
	\end{equation}
	where the error $(\delta_{x_i},\delta_{y_i})\sim\mathcal{N}^2\left(0,\sigma_{\mathrm{p},i}^2/2\right)$. 
	This position is periodically measured, and broadcast together with the variance $\sigma_{\mathrm{p},i}^2$ in a beacon signal. Upon the reception of such a beacon signal from $u_i$, every other $u_j\in\mathcal{U}$ nearby is not only informed about the position of $u_i$, but also able to estimate the distance $d_{i,j}(t)$ between them therefrom:
	\begin{equation}
			\tilde{d}_{i,j}=\left(\sqrt{\left({x}_j-{x}_i\right)^2+\left({y}_j-{y}_i\right)^2}+\delta_{d_{i,j}}\right)^+,
	\end{equation}
	where $x^+\triangleq\max\{0,x\}$. Depending on the specific mechanism of distance estimation, the error $\delta_{d_{i,j}}\sim\mathcal{N}\left(0,\sigma_{\mathrm{d},i}^2\right)$ may originate in the random channel fading, noise, interference, or jitters; and the variance $\sigma_{\mathrm{d},i}^2$ may be either consistent or random, either dependent on $d_i$ or not. Without loss of generality, in this work we assume that $\sigma_{\mathrm{d},i}^2$ is a random variable independent from $d_i$, and that it can be acquired by $u_0$ from the received beacon. Due to the limits in radio link coverage and speed of \glspl{uav}, the errors caused by mobility during the transmission latency of the beacon signals can be neglected. For convenience of analysis and discussion, in this study we focus on one arbitrary \ac{uav}, index it as $u_0$, and assume it is able to receive the beacon signals from all other $u_i$ where $i\in\{1,2,\dots I\}$. From here on, we simply notate $d_{i,0}$ as $d_i$. As illustrated in Fig.~\ref{fig:concept}, our aim is to implement an accurate estimator $\left[\hat{x}_0,\hat{y}_0\right]=\Phi\left(\tilde{x}_0,\tilde{y}_0, \mathbf{b}_1, \mathbf{b}_2\dots \mathbf{b}_I\right)$,
	where $\mathbf{b}_i=\left[\tilde{x}_i, \tilde{y}_i, \sigma_{\mathrm{p},i}^2, \tilde{d}_i, \sigma_{\mathrm{d},i}^2\right]$ is $u_i$'s beacon signal.
	
	\subsection{Attack Model}
	As discussed in Sec.~\ref{sec:introduction}, $J\leqslant I$ \glspl{uav} out of ${u_1,u_2,\dots u_I}$may be compromised by attacks (taking $u_0$'s point of view, itself is always trustworthy). In this study we consider four different attack modes as explained below.
	\subsubsection{Deterioration}
	In this mode, the attacker is able to increase the variances $\sigma_{\mathrm{p},i}^2$ and $\sigma_{\mathrm{d},i}^2$ for the compromised $u_i$, and $u_0$ is aware of this deterioration in the measurements of position and distance. Thus, the beacon signal received by $u_0$ can be modeled as
	 {\footnotesize $\check{\mathbf{b}}_i=\left[\tilde{x}_i+n_{x,i}, \tilde{y}_i+n_{y,i}, \sigma_{\mathrm{p},i}^2+\sigma_{\Delta\mathrm{p},i}^2, \left(\tilde{d}_i+n_{\mathrm{d},i}\right)^+, \sigma_{\mathrm{d},i}^2+\sigma_{\Delta\mathrm{d},i}^2\right]$}, 
	 where $\left(n_{x,i},n_{y,i}\right)\sim\mathcal{N}^2\left(0,\sigma_{\Delta\mathrm{p},i}^2/2\right)$ and $n_{\mathrm{d},i}\sim\mathcal{N}\left(0,\sigma_{\Delta\mathrm{d},i}^2\right)$. We refer to $\left[\sigma_{\Delta\mathrm{p},i}^2,\sigma_{\Delta\mathrm{d},i}^2\right]$ as the \ac{av}. Such effect can be achieved by jamming the beacon signals of $u_i$ and the \ac{gps}.
	
	\subsubsection{Variance}
	This mode is similar to the deterioration mode, except that it prevents $u_0$ from being aware of the introduced error. In other words, the received beacon can be modeled as
	$\check{\mathbf{b}}_i=\left[\tilde{x}_i+n_{x,i}, \tilde{y}_i+n_{y,i}, \sigma_{\mathrm{p},i}^2, \left(\tilde{d}_i+n_{\mathrm{d},i}\right)^+, \sigma_{\mathrm{d},i}^2\right]$, where $\left(n_{x,i},n_{y,i}\right)\sim\mathcal{N}^2\left(0,\sigma_{\Delta\mathrm{p},i}^2/2\right)$ and $n_{\mathrm{d},i}\sim\mathcal{N}\left(0,\sigma_{\Delta\mathrm{d},i}^2\right)$. Such effect can be commonly achieved either by tampering the original $\mathbf{b}_i$ through a man-in-the-middle attack, or by directly hijacking $u_i$ to falsify its position information.

	\subsubsection{Bias}
	In this mode, the attacker is able to bias the measured position $(x_i, y_i)$ without affecting the rest information in $\mathbf{b}_i$, so the beacon signal received by $u_0$ can be modeled as
	$\check{\mathbf{b}}_i=\left[\tilde{x}_i+\nu_{x_i}, \tilde{y}_i+\nu_{y_i}, \sigma_{\mathrm{p},i}^2, \tilde{d}_i, \sigma_{\mathrm{d},i}^2\right]$, where the \ac{av} $[\nu_{x_i}, \nu_{y_i}]$ remains constant over the time. Like the variance mode, it can be achieved either with a man-in-the-middle attack or by hijacking $u_i$.
	
	\subsubsection{Manipulation}
	This mode is an enhanced version of the bias mode. In addition to biasing the measured position, the attacker tampers the variances $\sigma_{\mathrm{p},i}^2$ and $\sigma_{\mathrm{d},i}^2$ to zero: $\check{\mathbf{b}}_i=\left[\tilde{x}_i+\nu_{x_i}, \tilde{y}_i+\nu_{y_i}, 0, \tilde{d}_i, 0\right]$.
	
	Motivated by the fact revealed in \cite{HKZ+2023trustawareness}, that a low penetration rate can make data-injection attacks more covert but yet effective, we allow the attacker(s) to commit attacks only at a random part of the beacon signals sent by the compromised \ac{uav}, by a penetration rate of $r\subscript{atk}$. This further leads us to define two different attacking schemes when more than one \glspl{uav} are compromised, namely
	
	\setcounter{subsubsection}{0}
	\subsubsection{Uncoordinated} where the attack is committed randomly and independently at every compromised \ac{uav}%
	; and
	\subsubsection{Coordinated} where the attack is always committed at all compromised \glspl{uav} simultaneously.

	\section{Robust Mutual Positioning}\label{sec:robust_positioning}
	\subsection{Least Square and Maximum Likelihood Estimators}\label{subsec:lse_mle}
	A straightforward and widely applied solution for multi-sensor positioning is the \ac{lse}:
	\begin{equation}\label{eq:lse}
		[\hat{x}_0,\hat{y}_0]=\arg\min\limits_{[x,y]}\sum\limits_{i\in\mathcal{I}}\left\vert\sqrt{\left(\tilde{x}_i-x\right)^2+\left(\tilde{y}_i-y\right)^2}-\tilde{d}_i\right\vert,
	\end{equation}
	where $\mathcal{I}=\{0,1,2\dots I\}$ and $\tilde{d}_0=0$.
	As a non-linear \ac{mmse} estimator it can be efficiently solved by the gradient descent algorithm \cite{GVW2012efficient}.
	
	The \ac{lse} is neglecting the difference in measurement accuracy among various \glspl{uav} $u_i$, which degrades its precision and  convergence efficiency when the measurements are significantly variant in quality. To address this issue, the \ac{mle} can be designed as
	\begin{equation}
		[\hat{x}_0,\hat{y}_0]=\arg\max\limits_{[x,y]}\mathcal{L}(\tilde{x}_0,\tilde{y}_0,\mathbf{b}_1,\mathbf{b}_2\dots\mathbf{b}_I\vert x,y).
	\end{equation}
	Leveraging the Bayesian estimation it can be transformed into
	\begin{equation}\label{eq:bayesian_mle}
		[\hat{x}_0,\hat{y}_0]=\arg\max\limits_{[x,y]}f_{x_0,y_0}(x,y~\vert~\tilde{x}_0,\tilde{y}_0,\mathbf{b}_1,\mathbf{b}_2\dots\mathbf{b}_I).
	\end{equation}
	Taking an approximation that $\sqrt{\left(\tilde{x}_i-x\right)^2+\left(\tilde{y}_i-y\right)^2}-\tilde{d}_i$ is a Gaussian variable $\varepsilon_i\sim\mathcal{N}(0,\sigma_i^2)$ for all $i\in\mathcal{I}$, \eqref{eq:bayesian_mle} is further transformed into a \ac{wlse}:
	\begin{equation}\label{eq:wlse}
		[\hat{x}_0,\hat{y}_0]=\arg\min\limits_{[x,y]}\sum\limits_{i\in\mathcal{I}}\frac{\left\vert\sqrt{\left(\tilde{x}_i-x\right)^2+\left(\tilde{y}_i-y\right)^2}-\tilde{d}_i\right\vert}{\sigma_i}.
	\end{equation}
	Especially, if $\sigma_i=\sigma$ for all $i\in\mathcal{I}$, it is equivalent to the \ac{lse} defined in Eq.~\eqref{eq:lse}.

	\subsection{Error Conversion}\label{subsec:err_conversion}
\begin{figure}[!htbp]
		\centering
		\includegraphics[width=.45\linewidth]{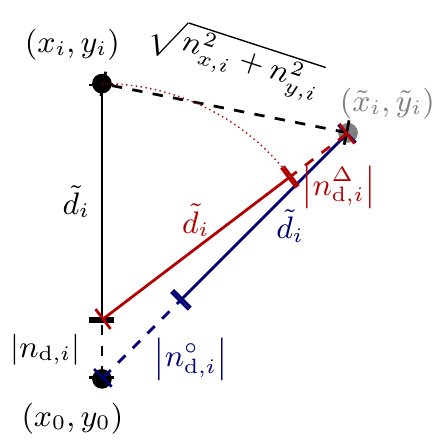}
		\caption{Error model mixing position and distance errors}
		\label{fig:err_conv}
	\end{figure}

	Now consider the error model shown in Fig.~\ref{fig:err_conv}. While the distance $\tilde{d}_i$ is measured from the actual position $(x_i,y_i)$ of the reporting $u_i$, the target \ac{uav} $u_0$ is only able to obtain its inaccurate position measurement $(\tilde{x}_i, \tilde{y}_i)$, and will consider the distance measurement as thereon based (the blue path in Fig.~\ref{fig:err_conv}). Thus, the effective distance error $n^\circ_{\mathrm{d},i}$ depends on not only $n_{\mathrm{d},i}$, but also an extra error $n^\Delta_{\mathrm{d},i}$ (the red dashed line).
	
	The distribution of $n^\Delta_{\mathrm{d},i}$ is as complex as analytically intractable (detailed derivation emitted in this paper due to the length limit). Nevertheless, it can be precisely approximated by a non-zero-mean Gaussian distribution $\mathcal{N}\left(\mu_i^\Delta,\left(\sigma_i^\Delta\right)^2\right)$. Both $\mu_i^\Delta$ and $\sigma_i^\Delta$ can be well fitted as paraboloid functions of $\tilde{d}_i$ and $\sigma_{\mathrm{p},i}$: a numerical fitting based on $1000$ random samples per combination $(\tilde{d}_i,\sigma_{\mathrm{p},i})$ is shown in Fig.~\ref{fig:err_conv}.
	\begin{figure}[!htpb]
		\centering
		\includegraphics[width=\linewidth]{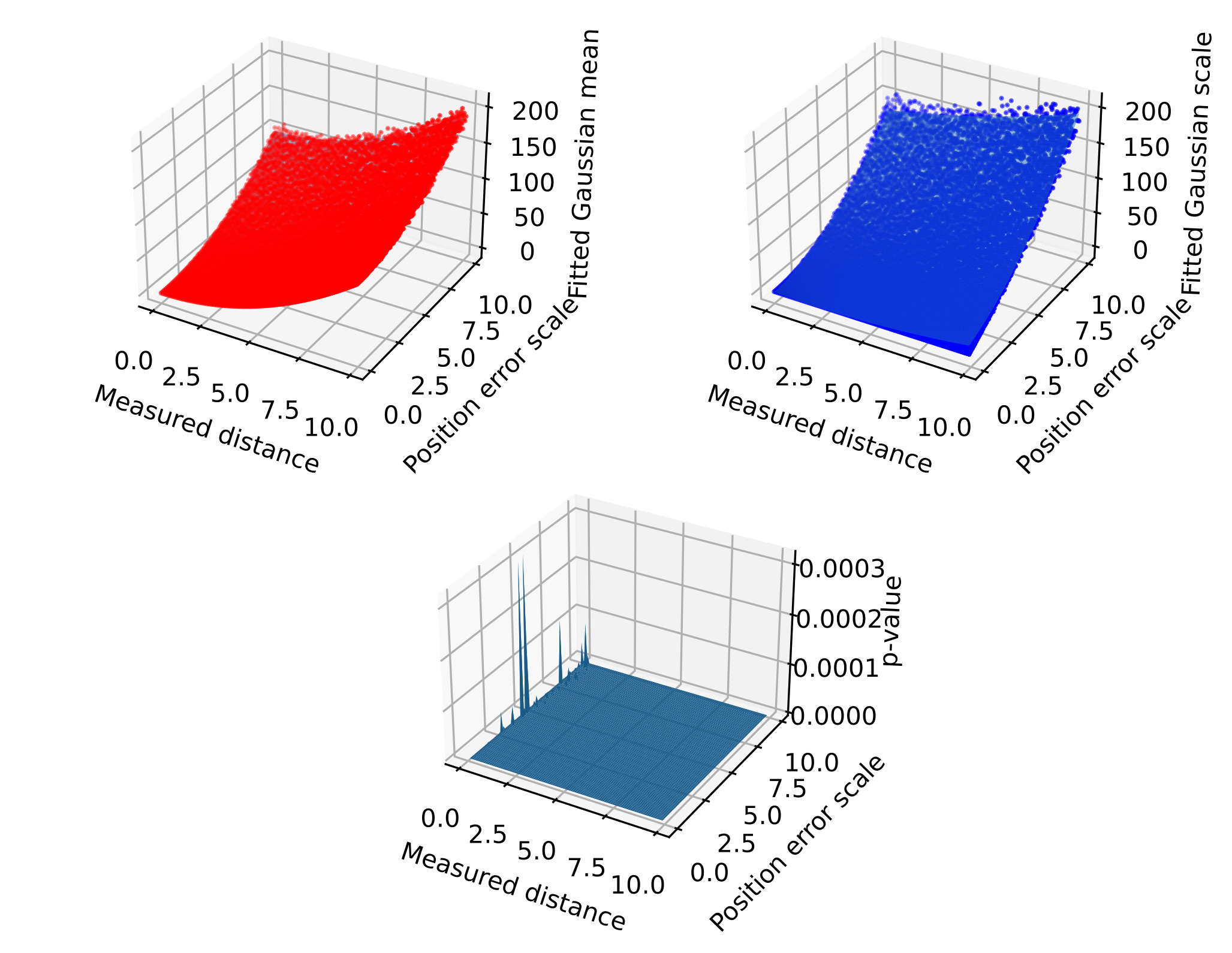}
		\caption{Paraboloid fitting of the Gaussian model for $n^\Delta_{\mathrm{d},i}$}
		\label{fig:err_conversion}
	\end{figure}
	Furthermore, since $\mu_i^\Delta$ and $\sigma_i^\Delta$ are independent from $\sigma_{\mathrm{d},i}$, we have $n^\circ_{\mathrm{d},i}\sim\mathcal{N}\left(\mu_i^\circ,\left(\sigma^\circ_{\mathrm{d},i}\right)^2\right)$ where $\mu_i^\circ=\mu_i^\Delta$ and $\left(\sigma^\circ_{\mathrm{d},i}\right)^2={\sigma_{\mathrm{d},i}}^2+\left(\sigma_i^\Delta\right)^2$. 
	
	\subsection{Robust Gradient Descent Algorithm}
	Since the overall noise can be Gaussian approximated, the \ac{wlse} in Eq.~\eqref{eq:wlse} can be still adopted under position errors as
	\begin{equation}
		\begin{split}
			&[\hat{x}_0,\hat{y}_0]\\
			=&\arg\min\limits_{[x,y]}\sum\limits_{i\in\mathcal{I}}\underset{\varphi_i(x,y)}{\underbrace{\frac{\left\vert\sqrt{\left(\tilde{x}_i-x\right)^2+\left(\tilde{y}_i-y\right)^2}-\tilde{d}_i+\mu^\circ_i\right\vert}{\sigma^\circ_i}}}.
		\end{split}
	\end{equation}
	And the gradient of the target function is 
	\begin{equation}
		\begin{split}
			&\nabla\left(\sum\limits_{i\in\mathcal{I}}\varphi_i\right)
			=\sum\limits_{i\in\mathcal{I}}\left[\frac{\partial\varphi_i}{\partial x},\frac{\partial\varphi_i}{\partial y}\right]\\
			=&[x,y]\sum\limits_{i\in\mathcal{I}}\frac{-2\mathrm{sgn}(\varphi_i)}{\sigma_i\sqrt{\left(\tilde{x}_i-x\right)^2+\left(\tilde{y}_i-y\right)^2}}.
		\end{split}
	\end{equation}
	Therewith, we can design the \ac{rgd} method as outlined in Algorithm~\ref{alg:rgd}, where the parameter $\epsilon$ sets the convergence floor, $K$ constrains the iteration number, $\alpha$ is the step length factor, $\gamma$ the step length discount when over-descended, and $m$ the momentum to stabilize the descending process. Especially, Line~\ref{line:err_conversion} applies the error model conversion to suppress the impact of position error, Line~\ref{line:weight_norm} normalizes the weight factors for different \glspl{uav}' measurement with respect to the least precise one, so as to keep the convergence speed on a reasonable level regardless of the overall measurement quality.
	\begin{algorithm}[!htpb]
		\caption{Robust Gradient Descent}
		\label{alg:rgd}
		\scriptsize
		\DontPrintSemicolon
		Specify: $\epsilon, K, \alpha, \gamma, m$\; 
		Input: $x\subscript{init}, y\subscript{init}, \mathcal{U},\left\{\mathbf{b}_i: \forall{u_i\in\mathcal{U}}\right\}$\;
		Initialize: $\hat{x}_0\gets x\subscript{init}, \hat{y}_0\gets y\subscript{init}, U_0\gets +\infty$\;
		\For(\tcp*[f]{Limit number of iterations}){$k\in\{1,2,\dots K\}$}{
			\ForEach{$u_i\in\mathcal{U}$}{
				Obtain $\left(\mu^\circ_i, \sigma^\circ_i\right)$ from $\mathbf{b}_i$\tcp*{Error conversion}\label{line:err_conversion}
				$D_i=\sqrt{\left(\tilde{x}_i-\hat{x}_0\right)^2+\left(\tilde{y}_i-\hat{y}_0\right)^2}$\label{line:Di}\;
				$s_i=\mathrm{sgn}\left(D_i-\tilde{d}_i+\mu^\circ_i\right)$\;
			}
			$\sigma\subscript{max}=\max\limits_{u_i\in\mathcal{U}}\sigma^\circ_i$\tcp*{Weight normalization}\label{line:weight_norm}
			$\Delta_x\gets m\hat{x}_0+\frac{\alpha}{\Vert\mathcal{U}\Vert_0}\sum\limits_{u_i\in\mathcal{U}}\frac{s_i\left(\tilde{x}_i-\hat{x}_0\right)\sigma\subscript{max}}{D_i\sigma^\circ_i}$\;
			$\Delta_y\gets m\hat{y}_0+ \frac{\alpha}{\Vert\mathcal{U}\Vert_0}\sum\limits_{u_i\in\mathcal{U}}\frac{s_i\left(\tilde{y}_i-\hat{y}_0\right)\sigma\subscript{max}}{D_i\sigma^\circ_i}$\;
			$\hat{x}_0\gets \hat{x}_0+\Delta_x, \hat{y}_0\gets \hat{y}_0+\Delta_y$\;
			$U_k\gets\frac{1}{\Vert\mathcal{U}\Vert_0}\sum\limits_{u_i\in\mathcal{U}}\frac{\left\vert D_i-\tilde{d}_i+\mu^\circ_i\right\vert\sigma\subscript{max}}{\sigma^\circ_i}$\;
			\uIf(\tcp*[f]{Over-descended}){$U_k>U_{k-1}$}{$\alpha=\alpha\gamma$\tcp*{Reduce step length}}
			\ElseIf(\tcp*[f]{Converged}){$\left(U_{k-1}-U_{k}\right)/U_{k-1}\leqslant\epsilon$}{\textbf{break}\;}
		}
		\Return{$\left(\hat{x}_0,\hat{y}_0, U_k\right)$}
	\end{algorithm}
	
	\section{Recursive Data Anomaly Detection}\label{sec:rdad}
	The \ac{rgd} algorithm can be vulnerable to attacks in two aspects: the attackers may \begin{enumerate*}[label=\emph{\arabic*)}]
		\item distort/falsify the position and/or distance measurements, so as to degrade the overall measurement quality; or
		\item falsify the reported variance of measurements, so as to misdirect the descent from the direction of gradient, or to destabilize the algorithm.
	\end{enumerate*}
	A secure solution is therefore in demand to \begin{enumerate*}[label=\emph{\roman*)}]
		\item protect the \ac{rgd} algorithm from false data; and
		\item identify abnormal data from normal ones.
	\end{enumerate*}
	
	To meet these requirements, we propose a two-stage recursive \ac{rdad} scheme as described in Alg.~\ref{alg:rdad}. The upper stage is basically executing the \ac{rgd} algorithm, one step by each iteration, but first calling the lower-stage recursion to identify and reject suspicious data before invoking \ac{rgd}. The lower stage is a confidence-based classification, given a reference position $\left(\hat{\hat{x}}_0,\hat{\hat{y}}_0\right)$ of $u_0$, it calculates the absolute error $\theta_i$ in distance measurement for every $u_i\in\mathcal{U}$ that $i\neq 0$, and estimates its cumulative density $\xi_i$  w.r.t. the converted mixed error model (Lines~\ref{line:absolute_err}--\ref{line:cdf}). Particularly, the converted error scale $\sigma^\circ_i$ is forced to be lower-bounded by a predetermined value $\sigma\subscript{min}$, in order to protect the algorithm from destabilization by extraordinarily low $\sigma^\circ_i$ (Line~\ref{line:err_bounding}). If $\xi_i$ exceeds a predetermined confidence degree $\Xi$, the measurement $\mathbf{b}_i$ will be classified as suspicious. In each iteration of the upper stage, the reference position is always reset to the local position measurement of target \ac{uav} $u_0$ (Line~\ref{line:anchor}) , for it is the only trustworthy anchor that is guaranteed unaltered for $u_0$ in potential existence of attackers. Then the lower-stage executes recursively, running for up to $L$ iterations,  in each iteration invoking the one-step \ac{rgd} to improve the reference position, and rejecting one most suspicious data, i.e. identifying it as anomaly. 
	
	\begin{algorithm}[!htpb]
		\caption{Recursive Data Anomaly Detection}
		\label{alg:rdad}
		\scriptsize
		\DontPrintSemicolon
		Specify: $\Xi,N, L,\sigma\subscript{min}, \epsilon, \alpha,\gamma,m$\;
		Input: $\mathcal{U},\left\{\mathbf{b}_i: \forall{u_i\in\mathcal{U}}\right\}$\; 
		Initialize: $\left(\hat{x}_0, \hat{y}_0\right)\gets\left(\tilde{x}_0, \tilde{y}_0\right), U_0\gets +\infty$\;
		\For(\tcp*[f]{Upper stage recursion}){$n\in\{1,2,\dots N\}$}{
			$\left(\hat{\hat{x}}_0,\hat{\hat{y}}_0\right)\gets\left(\tilde{x}_0, \tilde{y}_0\right)$\tcp*{Reset trustworthy local anchor}\label{line:anchor}
			\For(\tcp*[f]{Lower stage recursion}){$l\in\{1,2,\dots L\}$}{
				Update $\left(\hat{\hat{x}}_0,\hat{\hat{y}}_0\right)$ with Algorithm~\ref{alg:rgd}, $K = 1$\tcp*{One-step \ac{rgd}}
				\ForEach{$u_i\in\mathcal{U}\backslash\{u_0\}$}{
					Obtain $\left(\mu^\circ_i, \sigma^\circ_i\right)$ from $\mathbf{b}_i$\;
					$\sigma^\circ_i\gets\max\{\sigma^\circ_i,\sigma\subscript{min}\}$\tcp*{Counter manipulation}\label{line:err_bounding}
					$\theta_i = \left\vert\sqrt{\left(\tilde{x}_i-\hat{\hat{x}}_0\right)^2+\left(\tilde{y}_i-\hat{\hat{y}}_0\right)^2}-\tilde{d}_i+\mu^\circ_i\right\vert$\;\label{line:absolute_err}
					Compute the CDF $\xi_i=P_{\Theta_i}\left[\theta_i~\vert~\mu^\circ_i,\left(\sigma^\circ_i\right)^2\right]$\;\label{line:cdf}
				}
				$j\gets\arg\max\limits_{u_i\in\mathcal{U}}\xi_i$\;
				\uIf{$\xi_j>\Xi$}{
					Remove $u_i$ from $\mathcal{U}$\tcp*{Reject suspicious data}
				}
				\Else{
					\textbf{break}\;
				}
			}
			Update $\left(\hat{x}_0,\hat{y}_0, U_k\right)$ with Algorithm~\ref{alg:rgd}, $K = 1$ \tcp*{One-step RGD}
			\uIf(\tcp*[f]{Over-descended}){$U_k>U_{k-1}$}{$\alpha=\alpha\gamma$\tcp*{Reduce step length}}
			\ElseIf(\tcp*[f]{Converged}){$\left(U_{k-1}-U_{k}\right)/U_{k-1}\leqslant\epsilon$}{\textbf{break}\;}
		}
		\Return{$\left(\hat{x}_0,\hat{y}_0,\mathcal{U}\right)$}
	\end{algorithm}
	
	\section{Numerical Simulations}\label{sec:simulations}
	To demonstrate our analyses and evaluate our proposals, numerical simulations were conducted. We considered one target \ac{uav} and nine measuring \glspl{uav} that are randomly deployed w.r.t. a uniform distribution over a $\SI{30}{\meter}\times\SI{30}{\meter}$ map, and tested our solution as well as baseline methods regarding different scenarios of attacks and error models. Especially, when testing attacking scenarios, we considered three measuring \glspl{uav} to be jammed/hijacked at a penetration rate of $0.5$. More details of the setup are listed in Tab.~\ref{tab:setup}.
	
	\begin{table}[!htbp]
		\centering
		\caption{Simulation Setup}
		\label{tab:setup}
		\begin{tabular}{>{\cellcolor{white}}m{0.2cm} | m{1.5cm} l m{3.7cm}}
			\toprule[2px]
			&\textbf{Parameter}&\textbf{Value}&\textbf{Remark}\\
			\midrule[1px]
			
			\rowcolor{gray!30}
			& Map size &$\SI{30}{\meter}\times\SI{30}{\meter}$&\\
			&$I+1$&$10$& \# \glspl{uav}\\
			
			\rowcolor{gray!30}
			&$\sigma^2_{\mathrm{p},i}$&$\sim\mathcal{U}(0.1,2.1)$& position variance / \si{\meter^2}\\
			&$\sigma^2_{\mathrm{d},i}$&$\sim\mathcal{U}(0.1,0.9)$& distance variance / \si{\meter^2}\\
			
			\rowcolor{gray!30}
			\multirow{-5}{*}{\rotatebox{90}{\textbf{System}}} & $M$ & 1000 & \# Monte-Carlo tests per scenario\\
			\midrule[1px]
			
			&$J$ &$3$& \# compromised \glspl{uav} \\
			
			\rowcolor{gray!30}
			&$r\subscript{atk}$ & $0.5$&penetration rate\\
			&$\left[\sigma_{\Delta\mathrm{p},i}^2,\sigma_{\Delta\mathrm{d},i}^2\right]$ & $[\SI{50}{\meter^2},\SI{50}{\meter^2}]$&\ac{av}: deterioration \& variance\\
			
			\rowcolor{gray!30}
			\multirow{-4}{*}{\rotatebox{90}{\textbf{Attackers}}}&$\left[\nu_{x_i},\nu_{y_i}\right]$ & $[\SI{5}{\meter},\SI{5}{\meter}]$&\ac{av}: bias \& manipulation\\
			\midrule[1px]
			
			&$\epsilon$ & $1\times 10^{-6}\si{\meter}$&convergence floor\\
			
			\rowcolor{gray!30}
			&$K$ & $15$& max. iterations \ac{rgd}\\
			&$[N,L]$ & $[15,5]$& max. iterations \ac{rdad}\\
			
			\rowcolor{gray!30}
			&$\alpha$ & $0.9$ & initial step length factor\\
			&$\gamma$ & $0.9$ & step length discount\\
			
			\rowcolor{gray!30}
			&$m$ & $1\times 10^{-5}$ & momentum\\
			&$\Xi$ & $99\%$ & confidence degree\\
			
			\rowcolor{gray!30}
			\multirow{-8}{*}{\rotatebox{90}{\textbf{Positioning}}}&$\left(\sigma^\circ\subscript{min}\right)^2$ & $0.1\si{\meter^2}$ & lower bound of variance\\
			
			\bottomrule[2px]
		\end{tabular}
	\end{table}
	
	
	\subsection{Robustness Assessment}\label{subsec:robustness_assessment}
\begin{figure}[!htpb]
		\centering
		\begin{subfigure}{.85\linewidth}
			\centering
			\includegraphics[width=\linewidth]{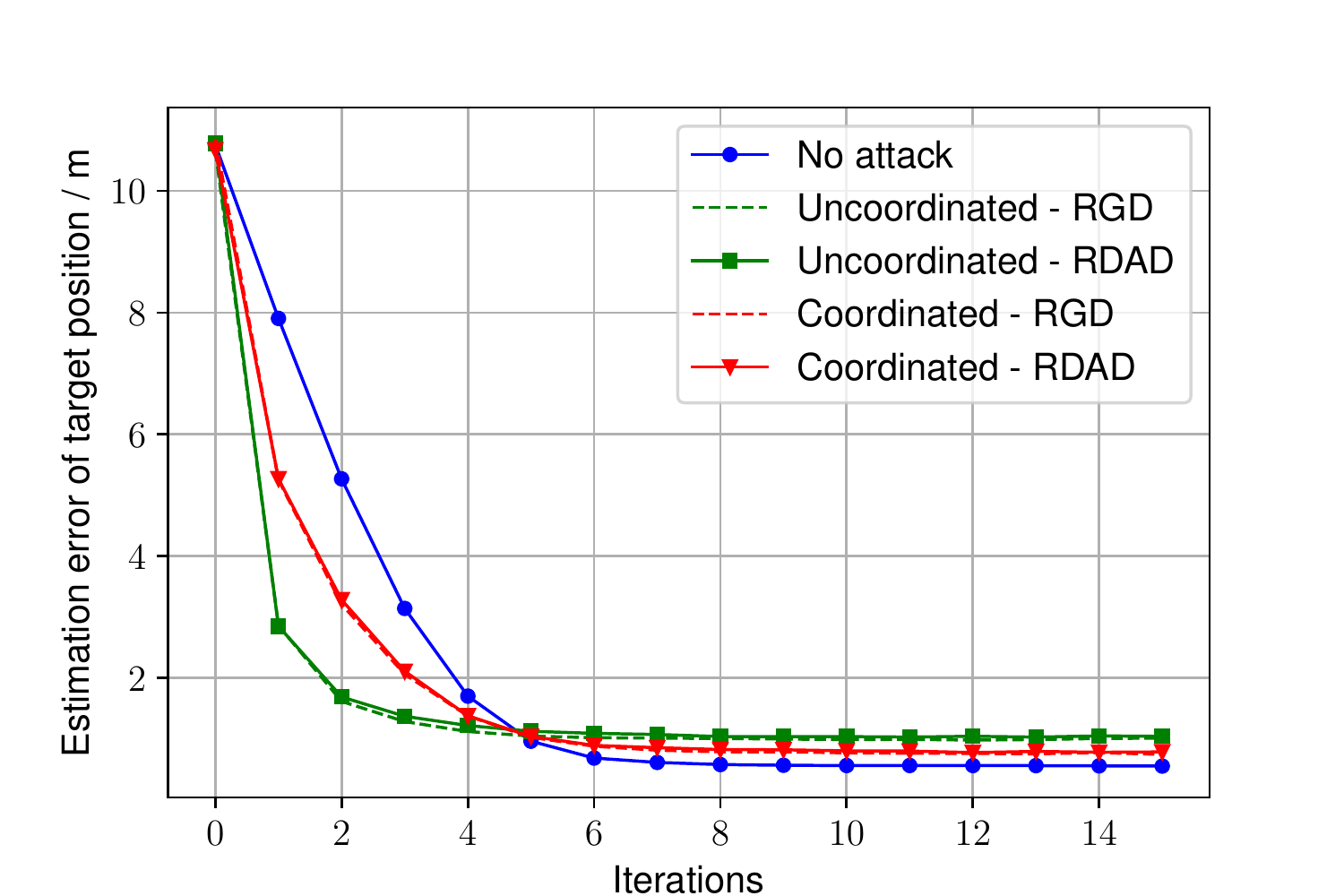}
			\subcaption{Attack mode: deterioration}
			\label{subfig:eval_deterioration}
		\end{subfigure}
		\begin{subfigure}{.85\linewidth}
			\centering
			\includegraphics[width=\linewidth]{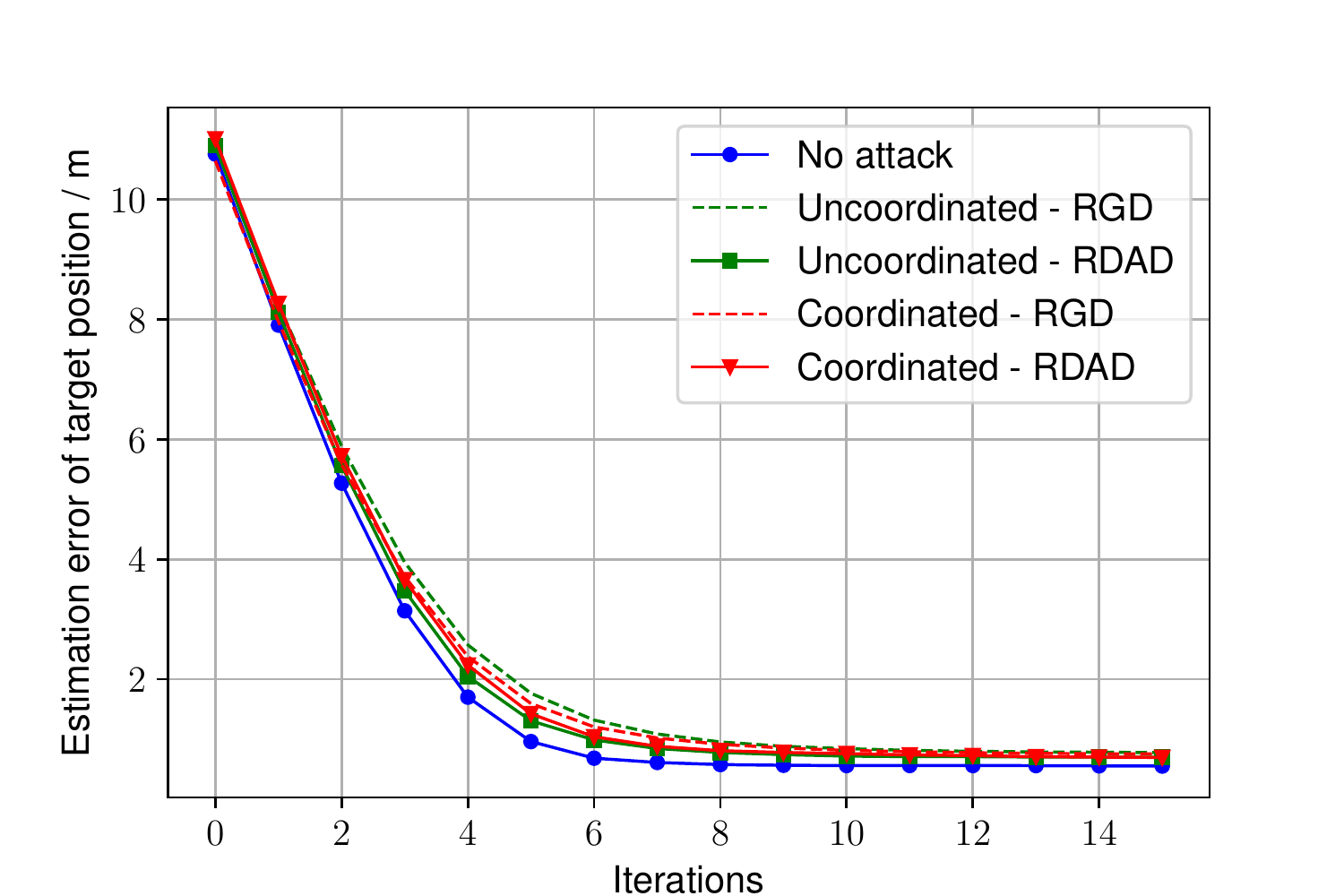}
			\subcaption{Attack mode: variance}
			\label{subfig:eval_variance}
		\end{subfigure}
		\\
		\begin{subfigure}{.85\linewidth}
			\centering
			\includegraphics[width=\linewidth]{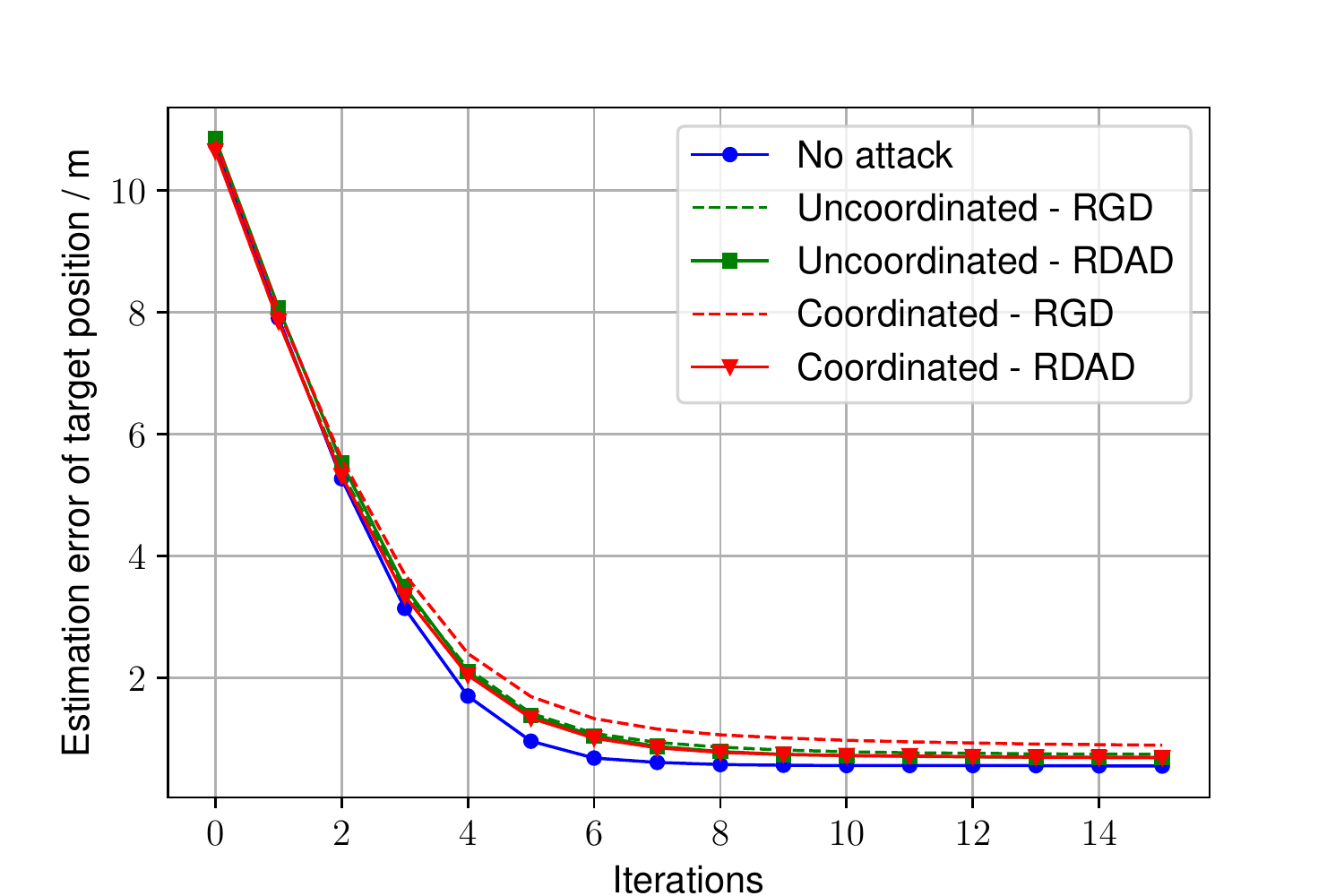}
			\subcaption{Attack mode: bias}
			\label{subfig:eval_bias}
		\end{subfigure}
		\begin{subfigure}{.85\linewidth}
			\centering
			\includegraphics[width=\linewidth]{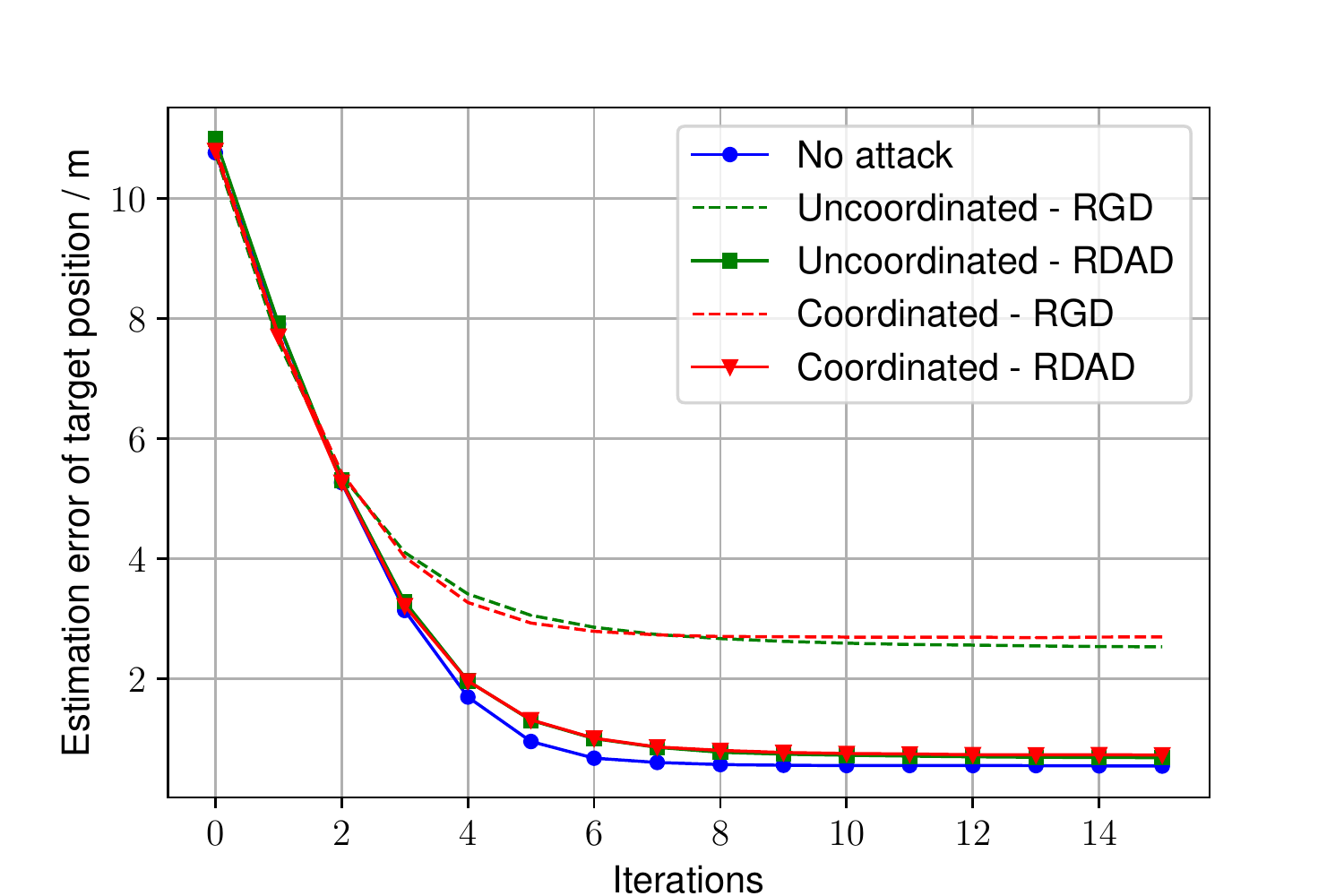}
			\subcaption{Attack mode: manipulation}
			\label{subfig:eval_manipulation}
		\end{subfigure}
		\caption{Convergence of the proposed methods under attacks}
		\label{fig:eval}
	\end{figure}

	We tested our \ac{rgd} algorithm under different attacking scenarios, both with and without the \ac{rdad} scheme. For the purpose of benchmark, we also tested \ac{rgd} in an attack-free scenario. The results are shown in Fig.~\ref{fig:eval}.
	
	In the scenario of deterioration attacks, as Fig.~\ref{subfig:eval_deterioration} shows, \ac{rgd} and \ac{rdad} are exhibiting similar performance, since it is hardly possible for the confidence-based \ac{rdad} scheme to identify compromised \glspl{uav} from normal ones with imprecise measurements, as they are honestly reporting the variance. Compared to the attack-free baseline, both methods converge faster when attacked, due to the significantly increased weights of normal \glspl{uav} in the \ac{rgd} procedure. But they end up with a slightly higher positioning error when converged, which is inevitably caused by the worsened overall data quality. Especially, such effects are more significant when the attacks at different \glspl{uav} are uncoordinated rather than coordinated.
	
	Against variance attacks, both \ac{rgd} and \ac{rdad} methods are similarly and sufficiently robust regarding the final accuracy, while \ac{rdad} is converging faster than \ac{rgd}. Moreover, variance attacks are more effective against \ac{rgd} when committed uncoordinated, but more effective against \ac{rdad} coordinated.
	
	Against bias attacks, \ac{rdad} shows a good robustness, rapidly converging to a high accuracy regardless of the coordination among compromised \glspl{uav}. \ac{rgd}, on the other hand, is working similarly well as \ac{rdad} under uncoordinated bias attacks, but significantly impacted by coordinated attacks.
	
	At last, when attacked in the manipulation mode, the accuracy of \ac{rgd} is dramatically degraded, with the error more than tripled in comparison to the attack-free baseline, regardless the coordination of attacks. To the contrary, \ac{rdad} effectively protects the system from attacks of this type.
	
	In summary, our proposed \ac{rgd} algorithm is robust against deterioration, variance, and uncoordinated bias attacks, but sensitive to coordinated bias attacks, and vulnerable to manipulation attacks. Its shortages, nevertheless, can be well resolved by our \ac{rdad} scheme.
	
	\subsection{\ACRshort{roc} Analysis of the \ac{rdad} Scheme}
\begin{figure}[!htpb]
		\centering
		\includegraphics[width=.9\linewidth]{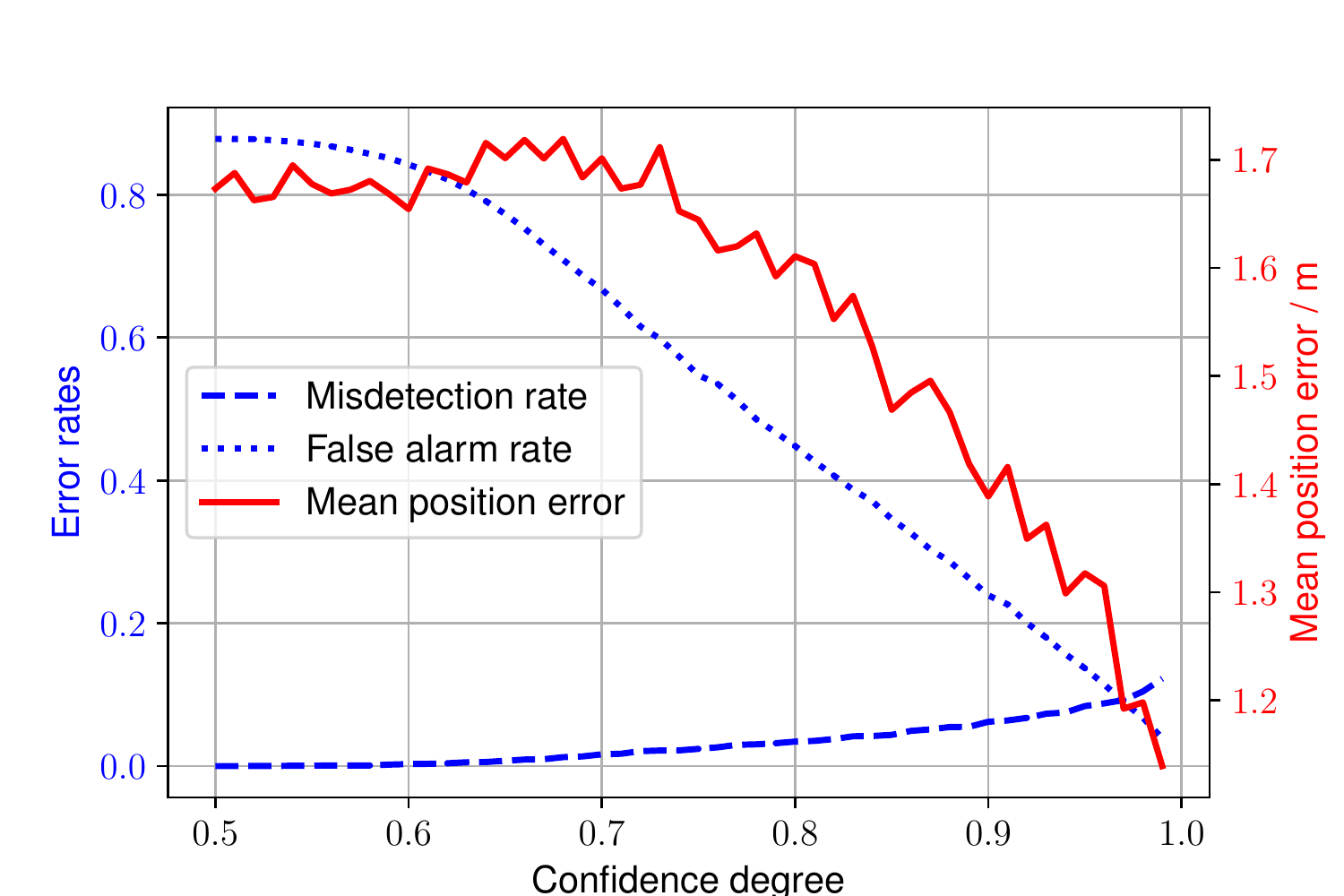}
		\caption{$\Xi$-sensitivity of \ac{rdad} against coordinated bias attacks}
		\label{fig:sensitivity_confidence}
	\end{figure}

	\ac{rdad} can be sensitive to the specification of confidence level $\Xi$, which determines its misdetection and false alarm rates in anomaly detection. As an example, the $Xi$-sensitivity of \ac{rdad} under coordinated bias attacks was evaluated through $10~000$ Monte-Carlo tests and the result is illustrated in Fig.~\ref{fig:sensitivity_confidence}. As the confidence degree increases, the algorithm becomes more tolerant to measurements with low likelihood, so that the false alarm rate drops, with a higher misdetection rate as the price. The overall positioning accuracy is therewith also influenced, but the impact is not monotonic of $\Xi$.
	
	The relation between the false alarm rate $r_\textrm{FA}$ and the misdetection rate $r_\textrm{MD}$ is the key feature of any detecting algorithm, commonly known as the \ac{roc}. The lower bound of all \ac{roc} curves is a straight line segment $r_\textrm{MD} = 1- r_\textrm{FA}$ over $r_\textrm{FA}\in[0,1]$,
	which implies a blind guess that randomly identifies data by ratio of $r_\textrm{FA}$ as anomalies, regardless their content. The \ac{roc} curve of an arbitrary rational detector must lie below this line, and the farther it distances from this bound, the better the detector is.
	
	\begin{figure}[!h]
		\centering
		\begin{subfigure}{\linewidth}
			\centering
			\includegraphics[width=.9\linewidth]{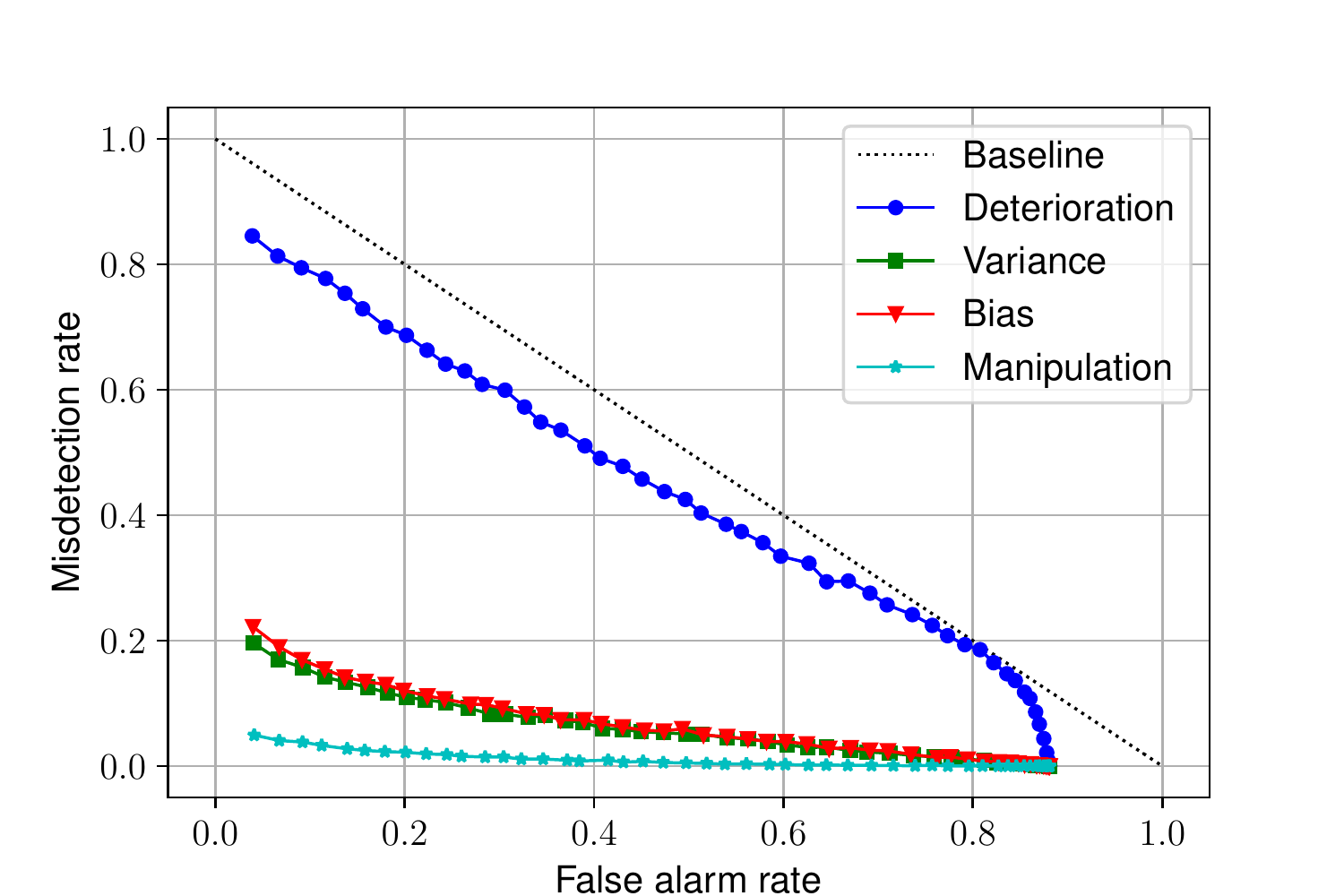}
			\subcaption{Uncoordinated attacks}
			\label{subfig:roc_uncoord}
		\end{subfigure}
		\begin{subfigure}{\linewidth}
			\centering
			\includegraphics[width=.9\linewidth]{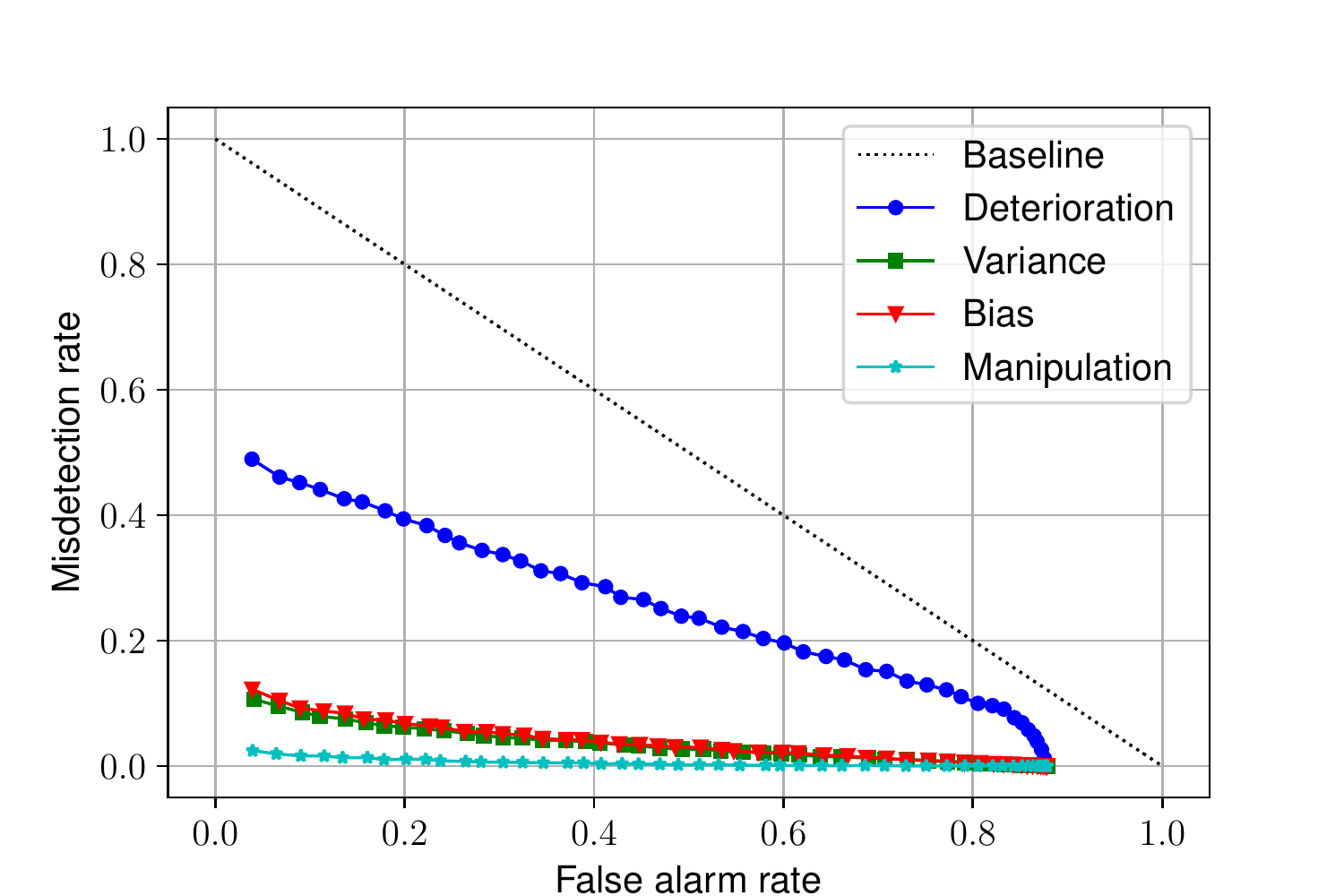}
			\subcaption{Coordinated attacks}
			\label{subfig:roc_coord}
		\end{subfigure}
		\caption{\Ac{roc} of \ac{rdad} against attacks of different modes.}
		\label{fig:roc}
	\end{figure}

	The results of \ac{roc} analysis to \ac{rdad} in different attack scenarios are shown in Fig.~\ref{fig:roc}. Once again, we see that our proposed approach exhibits a good performance in detecting attacks in modes of variance, bias, and manipulation. Especially, the manipulation attack, which we have in Sec.~\ref{subsec:robustness_assessment} demonstrated to be the most effective one on \ac{rgd}, can be the most accurately identified by \ac{rdad}. In contrast, deterioration attacks, like discussed in Sec.~\ref{subsec:robustness_assessment}, can be hardly detected due to its nature that cannot be distinguished from endogenous measurement errors. Furthermore, declining such deteriorated measurements with increased error will not be helpful but only destructive to the positioning accuracy, as long as they are neither biased or falsified in variance. Nevertheless, some use scenarios may still require a sensitive detection to such kind of attacks. This can be straightforwardly solved by applying a hard threshold on the measurement error scale $\sigma^\circ_i$, which is trivial and out of the scope of our work. In addition, the detection performance of \ac{rdad} is generally better against coordinated attacks than against uncoordinated ones, which agrees with the intuition that coordinated attacks are more exposed to the detector due to the significant pattern.

	\section{Conclusion and Outlooks}\label{sec:conclusion}
	In this paper, we have investigated the problem of multi-\ac{uav} mutual positioning, identified its use scenarios and technical challenges, and proposed a novel solution to address both the accuracy problem and the security risk. With numerical simulations, we have verified its robustness and detection accuracy against attacks of different modes.
	
	Yet there is one issue not addressed in this work but worth to remark, that due to the limited effective range of mutual radio positioning, compromised \glspl{uav} may locally outnumber normal ones in some regions and therewith break through the defense of our methods, even if they are only a minority in the global \ac{uav} set. Assessment of this risk and development of countering measures can be interesting for follow-up studies.

	\section*{Acknowledgment}
	This work is supported by the European Commission via the Horizon Europe project \emph{Hexa-X-II} (GA no. 101095759). B. Han (bin.han@rptu.de) is the corresponding author.



\bibliographystyle{IEEEtran}
\bibliography{references}

\end{document}

%% file: glossary.tex
\makeglossaries
\newacronym{6g}{6G}{Sixth Generation}
\newacronym{av}{AV}{attacking vector}
\newacronym{gps}{GPS}{Global Positioning System}
\newacronym{lse}{LSE}{least square estimator}
\newacronym{mec}{MEC}{Multi-Access Edge Computing}
\newacronym{mle}{MLE}{maximum likelihood estimator}
\newacronym{mmse}{MMSE}{minimum mean square error}
\newacronym{rdad}{RDAD}{Recursive Data Anomaly Detection}
\newacronym{rgd}{RGD}{Robust Gradient Descent}
\newacronym{roc}{ROC}{receiver operating characteristic}
\newacronym{wlse}{WLSE}{weighted \ac{lse}}
\newacronym{uav}{UAV}{unmanned aerial vehicle}

%% file: main.bbl
\begin{thebibliography}{10}
\providecommand{\url}[1]{#1}
\csname url@samestyle\endcsname
\providecommand{\newblock}{\relax}
\providecommand{\bibinfo}[2]{#2}
\providecommand{\BIBentrySTDinterwordspacing}{\spaceskip=0pt\relax}
\providecommand{\BIBentryALTinterwordstretchfactor}{4}
\providecommand{\BIBentryALTinterwordspacing}{\spaceskip=\fontdimen2\font plus
\BIBentryALTinterwordstretchfactor\fontdimen3\font minus
  \fontdimen4\font\relax}
\providecommand{\BIBforeignlanguage}[2]{{%
\expandafter\ifx\csname l@#1\endcsname\relax
\typeout{** WARNING: IEEEtran.bst: No hyphenation pattern has been}%
\typeout{** loaded for the language `#1'. Using the pattern for}%
\typeout{** the default language instead.}%
\else
\language=\csname l@#1\endcsname
\fi
#2}}
\providecommand{\BIBdecl}{\relax}
\BIBdecl

\bibitem{SSA+2019unmanned}
H.~Shakhatreh, A.~H. Sawalmeh \emph{et~al.}, ``Unmanned aerial vehicles
  ({UAVs}): {A} survey on civil applications and key research challenges,''
  \emph{IEEE Access}, vol.~7, pp. 48\,572--48\,634, 2019.

\bibitem{SAB+2019design}
R.~Shakeri, M.~A. Al-Garadi \emph{et~al.}, ``Design challenges of multi-{UAV}
  systems in cyber-physical applications: {A} comprehensive survey and future
  directions,'' \emph{IEEE Commun. Surv. Tutor.}, vol.~21, no.~4, pp.
  3340--3385, 2019.

\bibitem{JHH+2021road}
W.~Jiang, B.~Han \emph{et~al.}, ``The road towards {6G}: {A} comprehensive
  survey,'' \emph{IEEE Open J. Commun. Soc.}, vol.~2, pp. 334--366, 2021.

\bibitem{LCG+2022multi}
Z.~Liu, Y.~Cao \emph{et~al.}, ``Multi-{UAV} network assisted intelligent edge
  computing: {Challenges} and opportunities,'' \emph{China Commun.}, vol.~19,
  no.~3, pp. 258--278, 2022.

\bibitem{AMH2016rssi}
P.~Abouzar, D.~G. Michelson \emph{et~al.}, ``{RSSI}-based distributed
  self-localization for wireless sensor networks used in precision
  agriculture,'' \emph{IEEE Trans. Wirel. Commun.}, vol.~15, no.~10, pp.
  6638--6650, 2016.

\bibitem{YS2015wifi}
C.~Yang and H.-r. Shao, ``{WiFi}-based indoor positioning,'' \emph{IEEE Commun.
  Mag.}, vol.~53, no.~3, pp. 150--157, 2015.

\bibitem{DSM+2021positioning}
S.~Dwivedi, R.~Shreevastav \emph{et~al.}, ``Positioning in {5G} networks,''
  \emph{IEEE Commun. Mag.}, vol.~59, no.~11, pp. 38--44, 2021.

\bibitem{KZS+2020review}
T.~Kim~Geok, K.~Zar~Aung \emph{et~al.}, ``Review of indoor positioning: Radio
  wave technology,'' \emph{Appl. Sci.}, vol.~11, no.~1, p. 279, 2020.

\bibitem{MMS+2013navigation}
T.~Maki, T.~Matsuda \emph{et~al.}, ``Navigation method for underwater vehicles
  based on mutual acoustical positioning with a single seafloor station,''
  \emph{IEEE J. Ocean.}, vol.~38, no.~1, pp. 167--177, 2013.

\bibitem{WWS+2020auxiliary}
F.~Wen, J.~Wang \emph{et~al.}, ``Auxiliary vehicle positioning based on robust
  {DOA} estimation with unknown mutual coupling,'' \emph{IEEE Internet Things
  J.}, vol.~7, no.~6, pp. 5521--5532, 2020.

\bibitem{WB2019robust}
J.~Won and E.~Bertino, ``Robust sensor localization against known sensor
  position attacks,'' \emph{IEEE Trans. Mob. Comput.}, vol.~18, no.~12, pp.
  2954--2967, 2019.

\bibitem{GVW2012efficient}
R.~Garg, A.~L. Varna \emph{et~al.}, ``An efficient gradient descent approach to
  secure localization in resource constrained wireless sensor networks,''
  \emph{IEEE Trans. Inf. Forensics Secur.}, vol.~7, no.~2, pp. 717--730, 2012.

\bibitem{MSK2021rss}
B.~Mukhopadhyay, S.~Srirangarajan \emph{et~al.}, ``{RSS}-based localization in
  the presence of malicious nodes in sensor networks,'' \emph{IEEE Trans.
  Instrum. Meas.}, vol.~70, pp. 1--16, 2021.

\bibitem{HKZ+2023trustawareness}
B.~Han, D.~Krummacker \emph{et~al.}, ``Trust-awareness to secure swarm
  intelligence from data injection attack,'' May 2023, to appear at IEEE ICC
  2023, preprint available at arXiv:2211.08407.

\end{thebibliography}
